\documentclass[10pt]{article}
\usepackage{graphicx}
\usepackage{amssymb,amsmath}
\usepackage{float}
\usepackage{tcolorbox}
\usepackage{amsthm}
\usepackage{graphicx}
\usepackage{hyperref}
\hypersetup{
	colorlinks=true,
	linkcolor=blue,
	filecolor=magenta,      
	urlcolor=cyan}

\usepackage[title]{appendix}
\usepackage{multirow}

\begin{document}

\title{Circle packing on spherical caps}
\author{Paolo Amore \\
\small Facultad de Ciencias, CUICBAS, Universidad de Colima,\\
\small Bernal D\'{i}az del Castillo 340, Colima, Colima, Mexico \\
\small paolo@ucol.mx}

\maketitle

\begin{abstract}
We have studied the packing of congruent disks on a spherical cap, for caps of different size and number of disks, $N$. 
This problem has been considered before only in  the limit cases of circle packing inside a circle  and on a sphere (Tammes problem), whereas all intermediate cases are unexplored. Finding the preferred packing configurations for a domain with both curvature and border could be useful in the description of physical and biological systems (for example, colloidal suspensions or the compound eye of an insect), with potential applications in engineering and architecture (e.g. geodesic domes).

We have carried out an extensive search for the densest packing configurations of congruent disks on spherical caps of selected angular widths 
($\theta_{\rm max} = \pi/6$, $\pi/4$, $\pi/2$, $3\pi/4$, $5 \pi/6$) and for several values of $N$. The
numerical results obtained in the present work have been used to establish (at least qualitatively) some general features for these configurations, in particular the behavior of the packing fraction as function of the number of disks and of the angular width of the cap, or the nature of the topological defects in these configurations ({  it was found that as the curvature increases, the overall topological charge on the border tends to become more negative}). 

Finally we have studied the packing configurations for $N = 19$, $37$, $61$, $91$ (hexagonal numbers) for caps  ranging from the flat disk to the whole sphere, to observe the evolution (and eventual disappearance) of the curved hexagonal packing configurations while increasing  the curvature.
\end{abstract}

\maketitle

\section{Introduction}
\label{sec:intro}

In this paper we study the dense packing configurations of a number of congruent disks disposed on a spherical cap. 
In the infinite plane the densest packing corresponds to a configuration where each disk is at the center of a regular hexagon, with its vertices occupied by congruent disks (the packing fraction is $\rho_{\rm plane} =\pi/\sqrt{12}$). However, if the domain is both curved and with a border, perfect hexagonal packing is impossible.

It is well known, in fact, that a sphere cannot be tiled only with hexagons, a consequence of Euler's theorem of topology. Additionally, the border introduces geometrical frustration even in the absence of curvature and generally makes it impossible to have a hexagonal packing. Only for certain domains, such as the equilateral triangle and the regular hexagon, and for specific numbers of disks, is it possible to have a perfect tiling of the container. However, the density achieved is nonetheless less than $\rho_{\rm plane}$ because of the area being wasted close to the border.

For the problem that we are considering in this paper the relative effect of the curvature and of the border can be
tuned in terms of the parameter $\delta = \mathcal{A}_{\rm cap}/\mathcal{A}_{\rm sphere}$ ($\mathcal{A}_{\rm cap}$ is area of the spherical cap and $\mathcal{A}_{\rm sphere}$ is the area of the sphere). 
In fact the effect of the curvature becomes negligible for $\delta \rightarrow 0$  and essentially one deals with circle packing inside a circle~\cite{Reis75,Melissen94b,Graham98,Graham97b,Fodor99,Fodor00,Fodor03,Nemeth98}, whereas the border disappears for $\delta \rightarrow 1$ (it shrinks down to a point) and one is left with circle packing on a sphere~\cite{Tammes30,Clare86,Clare91, Kottwitz91, Glover23}. The latter is also known as {\sl Tammes problem} in honor of the researcher that first introduced the problem~\cite{Tammes30}. Interestingly, Tammes was neither a mathematician nor a physicist, but a botanist, who was interested in describing the distribution of pores on pollen grains. The generalization of Tammes problem to higher dimensions, essentially packing on hyper--spheres, is being actively pursued in mathematics; the optimal configurations for this problem go under the name of {\sl spherical codes} (see for example \cite{Sloane81,Sloane84,Sloane13}). Bounds on codes in spherical caps have been obtained in refs.~\cite{Barg07,Bachoc09}.
A different problem, still regarding packing on a finite domain on a sphere, has been studied by Melissen~\cite{Melissen98} who considered the packing of up to $15$ caps on a spherical octant. 
Numerical results and plots for packing configurations corresponding to different geometries and dimensions are available at the repositories \cite{FriedmanRepo, SpechtRepo, SloanRepo, CohnRepo}: in particular the reader will find (among many other problems) in ref.~\cite{SpechtRepo} configurations for packing inside a circle in 2d and in refs.~\cite{SloanRepo, CohnRepo} for packing on a sphere in 3d (Tammes problem).

Although the two limiting cases (packing inside a circle and packing on a sphere) have been studied earlier, it appears that the general problem 
that we are considering in the present paper has not, with the notable exception of ref.~\cite{Appelbaum99}, where the authors attempted to determine nearly optimal packing configurations of up to $40$ disks on a hemisphere (this search however was empirical and limited to few disks, assuming a specific ansatz for the solutions). 
The motivation of ref.~\cite{Appelbaum99} was constructing a device to measure solar radiation by disposing up to $40$ detectors on an hemisphere, in a way that the area covered by the detectors could be as large as possible (this device has then been built as reported in ref.~\cite{Appelbaum20}).   

{  A  second exception is ref.~\cite{Tarnai15}, where solutions for $N=2,3,4,5,6,7,8$ have been conjectured on the basis of  a computer analysis.}

While the discrete problem that we have stated has not been studied for caps of arbitrary size, there has been a lot of effort in developing a description of how defects form and distribute on flat and curved surfaces, in the limit of a continuous elastic medium~\cite{Bowick02,Bowick07,Bowick13,Grason14,Brojan15,Grason16,Grason16b,Grason19,Travesset19,Giomi20,Hilgenfeldt21}. 
Additionally, Saff and collaborators have obtained asymptotic formulas for the  energy of systems of repelling charges on spheres in the limit  $N \rightarrow \infty$ ($N$ is the number of charges), for different Coulomb--like potentials~\cite{Saff94,Saff94b,Saff97,Saff19}. 

An in--depth numerical search has been conducted for the case of point charges constrained on a sphere~\cite{Wales06,Wales09} (Thomson problem) or on a surface of constant mean curvature~\cite{Wales13}, with the aid of the basin--hopping (BH) method, possibly the most effective method for the search of global minima~\cite{Li87,Wales97,Wales99,Wales03}. 
Qingyou and Grason\cite{Grason21} have aso performed a comparative study of discrete and continuum models of conformal crystals (the charges are interacting with a Coulomb potential and are subject in this case to a harmonic or anharmonic confinement, and not to a hard wall). All these works, however, deal with Coulomb--like repulsive interactions and look for configurations of minimal energy. In our case, the interaction is essentially a contact interaction between rigid, congruent caps and the quantity that is being optimized is the {\sl packing fraction} (i.e. the ratio between the total area covered by the disks and the area of the domain). A further difference comes from the topology of the problem (spherical cap), which has not been considered earlier for packing problems. 

Exploring the behavior of topological defects that emerge in the  packing  on a spherical cap may be useful for the description  of certain physical and biological systems, such as in the evaporation/drying of colloidal droplets~\cite{Manoharan03,Lauga04,Monteux11,Marin11,Marin12,Manoharan15,Mukhopadhyay21,Fan22}, in colloidal 
crystals~\cite{Vitelli10,Guerra18,Das22}, in the assembly of viral capsids~\cite{Zandi04,Twarock04,Rochal17} or 
in the  morphology of insect's eye~\cite{Kim16,Hayashi22}.


{ One should recognize that an insect's eye is a highly efficient, Nature-made detector, where the close packing of sensors on a curved surface is the result of  a long evolutive process, not of a computational algorithm.}   As a matter of fact there has been a consistent effort in recent years in developing "bio-inspired" hemispherical cameras, drawing inspiration from the insect's eyes~\cite{Leblebici13,Song13,Cogal16,Guenter17,Shi17,Lee18,Kim20,Blair21}.
{ Finally,  another application of circle packing on a spherical cap is in the design of geodesic domes for engineering and architecture~\cite{Tarnai11,Gaspar22,Bysiec23}. A key reference on this topic is the book by Popko~\cite{Popko21}.}

It is also important to recognize the multidisciplinary nature of packing problems, which are relevant in many different areas (mathematics, physics, biophysics, computer science, etc). In particular it is remarkable that one can use real physical/biological systems to shed light on difficult mathematical aspects of packing problems: one example of this is the "sausage conjecture" of Fejes-Toth, which concerns the shape of a convex hull with minimal of sphere packing in $d$ dimensions.
For $d = 2$ the "discs (2-dimensional balls) must be arranged so as to make their convex hull, as similar to a regular hexagon as possible"~\cite{Fejes-Toth75},
whereas for large dimensions ($d \geq 5$) it becomes convenient to distribute the objects in a linear configuration. The intermediate case, $d=3,4$, 
the problem is very difficult, or in Fejes-Toth's words, the "problem must be considered for $n = 3$
and $4$ as hopeless". Quite recently the authors of \cite{Marin23} have tackled this problem experimentally by studying arrangements of colloids in a ﬂaccid lipid vesicle, identifying both  linear, planar, and cluster conformations of spheres.

The main goal of this paper is to perform a numerical exploration of circle packing on spherical caps of different angular width, for a wide range of $N$ (number of disks). In this numerical exploration we want to study the properties of the topological defects emerging in the domain and establish through empirical observation the behavior of these systems, both as a function of the curvature and of the number of constituents $N$.  Moreover, for special values of $N$, corresponding to so--called "hexagonal numbers" ($N=3k (k+1)+1$ with $k=1,2,\dots$), Graham and Lubachevsky~\cite{Graham97b} discovered long time ago that for $N \leq 91$ ($k \leq 5$), the densest packing configurations in the circles are highly symmetrical and invariant under rotations of $\pi/3$ (in fact they are analogous to the hexagonal configurations found in an hexagon at these values of $N$). These authors called such configurations "curve hexagonal packing" (CHP) (recently the same kind of configurations have been found in regular polygons with a number of sides multiple of $6$ in ref.~\cite{Amore23b}). As the curvature of the spherical cap increases, at these values of $N$, the CHPs are modified and eventually bound to disappear. It is easy to figure out the fate of CHPs, by taking into account that the density of CHP is completely determined by the requirement that the border be completely packed: for larger curvature there is additional internal space available for packing which cannot be used by the CHP, since the size of the disks is already fixed. At some critical value of the angular width of the cap, this extra space may suddenly be used to produce denser, non--symmetrical configurations, by removing disks from  the border and populating the interior of the circle. We have carried out detailed numerical experiments for these configurations up to $N=91$ ($k=5$). 

The paper is organized as follows: in section~\ref{sec:method} we outline the general problem and 
describe the computational method for finding packing configurations on a spherical cap; in section ~\ref{sec:res} we present the numerical results obtained with our method and provide a detailed analysis of these results; in section \ref{sec:CHP} we discuss the behavior of CHP when $\theta_{\rm max}$ is being changed;
finally, in section ~\ref{sec:concl}  we draw our conclusion and identify possible directions for future research.

\section{The method}
\label{sec:method}

Before dealing with the specifics of our problem it is useful to describe briefly the packing algorithm that we use to look for dense configurations.

The starting point  is the algorithm originally devised by Nurmela and \"Ostergard in ref.~\cite{Nurmela97} for generating the packing of congruent disks inside 
a square (and later applied to the circle by Graham et al. in \cite{Graham98}) and the modified algorithms recently 
introduced by myself and collaborators in refs.~\cite{Amore23a, Amore23b, Amore23c, Amore23d}. 
In all these approaches the original optimization problem is transformed into an optimization problem of equal size charges 
inside a domain, repelling with a Coulomb-like interaction
\begin{equation}
V_{ij} = \left(\frac{\lambda}{r^2_{ij}} \right)^s  \ , 
\label{eq_V}
\end{equation}
where $r_{ij}$ is the euclidean distance between the two charges, $s > 0$ is a parameter that controls the range of the interaction and $\lambda >0$ is the minimal squared distance between any two charges. 
The charges correspond to the centers of the disks that are being packed in the domain. By parametrizing the coordinates of the charges in a suitable way, as done for example in ref.~\cite{Nurmela97} for the square, or in ref.~\cite{Amore23a} for regular polygons, the original constrained optimization problem is transformed into an unconstrained one (which is easier to deal with). 

At first, the coordinates of the charges are chosen randomly in the domain, and the charges are left to repel each other with a sufficiently long--range interaction ($s \approx 5-10$). Under the effect of the interaction the configuration of charges reaches the equilibrium, corresponding to an ordered arrangement of points on the domain. The procedure is then repeated by progressively increasing the value of $s$ (i.e. making the range of the interaction shorter and shorter) while updating $\lambda$, until reaching values of $s$ ($s \geq 10^8$) which correspond essentially to a contact interaction ($V = \infty$ for $r^2 \leq \lambda$ and $V=0$ for $r^2>\lambda$). The final points obtained at the end are then promoted to be the center of disks of radius $\sqrt{\lambda^{(\rm fin)}}/2$ and the packing fraction (density) is obtained as the ratio between the area covered by the disks and the area of the domain. The algorithm does not provide a direct optimization of the packing fraction and  this procedure has to be repeated many times in order to find denser configurations.


{The primary limitation of the original algorithm presented in Ref.\cite{Nurmela97}, as well as the algorithms discussed in Refs.\cite{Amore23a, Amore23b, Amore23c}, is that they are restricted to specific geometrical domains, such as squares, circles, or regular polygons. This limitation arises because the algorithm does not work directly with the physical container but with a smaller domain. Consequently, a scale transformation must be applied to relate the two domains, otherwise the nature of the problem would change during the process, rendering any comparison between configurations meaningless.}

In ref.~\cite{Amore23d}, however, this problem has been completely solved by introducing in the problem auxiliary degrees of freedom (image charges) that allow one to work directly with the physical domain. Just like in classical electrostatics, an image charge is a fictitious charge placed outside the domain on a line joining the physical charge that crosses the border normally. Each physical charge may have more than one image, although only the ones that are closer to the border are relevant. While image charges do not interact between themselves (they can even get very close to each other) they interact with the physical charges with the same interaction of eq.~(\ref{eq_V}), with the only difference that now $\lambda$ must also take into account the squared distance between charges and image charges. If a physical charge gets too close to the border of the domain it will feel a strong repulsion from its image charge and it will be pushed inwards. In this way, when $s$ is very large, all the physical charges will correspond to the centers of well--packed disks inside the domain, as desired.

As we proved in ref.~\cite{Amore23d} this simple recipe is quite successful and it opens the doors to wide class of problems that could not be dealt with using the original algorithm: in fact the cases studied in ref.~\cite{Amore23d} include rectangles, crosses, ellipses, the symmetric annulus and even the cardioid. Notably in this list we find domains that are concave, multiply connected and even with border singularities (the cardioid).

The domain that we are considering in this paper, i.e. a spherical cap in three dimensions, is another example where the 
basic algorithm of ref.~\cite{Nurmela97} or the versions of refs.~\cite{Amore23a, Amore23b, Amore23c}, cannot be applied because of the
lack of scale invariance of the domain. Also in this case one can apply the algorithm of ref.~\cite{Amore23d}  to 
work with a fixed physical domain with a modest increase in the computational complexity, due to the extra degrees of freedom in the problem (for a spherical cap, however, the computation of the image charges is simpler than in more complex geometries).

\begin{figure}[t]
\begin{center}
\bigskip\bigskip\bigskip
\includegraphics[width=7cm]{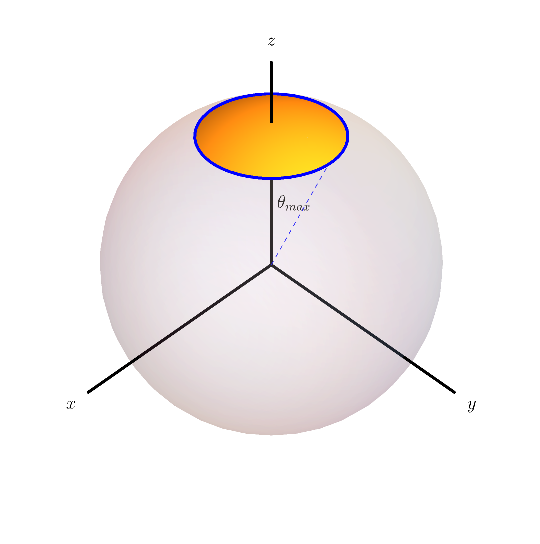}
\bigskip
\caption{A spherical cap of angular width $\theta_{\rm max}$.}
\label{Fig_cap}
\end{center}
\end{figure}

Fig.~\ref{Fig_cap}  provides a schematic representation of the domain that we are considering in this work, a spherical cap with angular width $\theta_{\rm max}$.  We find convenient to work with spherical caps of area $\pi$ (the area of the unit disk) and therefore the spherical cap lies on a sphere of radius $R = (2 \sin \left(\frac{\theta_{max} }{2}\right))^{-1}$.
The area of the sphere is then $\mathcal{A}_{\rm sphere} =4 \pi R^2 = \pi/\sin^2\left(\frac{\theta_{\rm max}}{2}\right) $ and  consequently $\delta = \sin^2\left(\frac{\theta_{\rm max}}{2}\right)$. 

A point on the spherical cap is identified in terms of two angles, $\theta$ and $\phi$ and the image  point that is closest to this point is found at an angle $2 \theta_{\rm max} - \theta$. As usual it is convenient to use a parametrization that converts the problem to an unconstrained optimization problem: this is done by expressing $\theta$  as 
\begin{equation}
\theta(t) = \theta_{\rm max} \sin^2 t , 
\end{equation}
where $t$ is a real parameter. As a result the condition $0 \leq \theta \leq \theta_{\rm max}$ is automatically enforced in the problem. After expressing the full energy functional, which includes the extra dofs (refer to eqs.(1) and (3) of ref.~\cite{Amore23d}), the problem can be directly attacked along the lines explained in our previous papers, in particular refs.~\cite{Amore23a} and  \cite{Amore23d} (the reader will find a pseudocode of the basic algorithm in the appendix of \cite{Amore23d})

Once a configuration is obtained numerically, its packing fraction  can be expressed in terms of $\theta_{\rm max}$ (the angular width of the cap) and of $\theta_{\rm min}$ (the angular width of each disk on the cap) as:
\begin{equation}
\rho = N \frac{\sin^2 \frac{\theta_{\rm min}}{2}}{\sin^2 \frac{\theta_{\rm max}}{2}} , 
\label{eq_dens}
\end{equation}
with
\begin{equation}
\cos 2\theta_{\rm min} =  \max_{i \neq j} \left[ {\bf r}_i \cdot {\bf r_j}\right] \ . 
\end{equation}

Because the number of local maxima of the density increases strongly with $N$ (number of points or equivalently number of disks), finding the global maximum of the packing fraction is extremely challenging (for the simpler case of the Thomson problem, for example, the number of independent configurations  grows exponentially~\cite{Erber91,Calef15}). Therefore finding good candidates for the global maxima of the density requires a thorough exploration of the density landscape, which becomes increasingly demanding for larger $N$: for example, Wales and collaborators~\cite{Wales06,Wales09} have applied the basin-hopping algorithm~\cite{Li87,Wales97,Wales99,Wales03} to the Thomson problem, improving several previous records and finding good candidates for global minima of the energy. The Tammes problem (circle packing on a sphere) differs from Thomson's problem because it requires maximizing the packing fraction, whereas the latter looks for a minimum of the total electrostatic energy. In this case the optimization needs to be  performed indirectly, 
increasing the computational complexity of the problem. 

Using algorithms for global optimization is however important when dealing with packing problems due to their inherent complexity.
Addis et al., for example, have used variants of the BH algorithm to perform circle packing in a square~\cite{Locatelli08}, while more recently Lai and collaborators~\cite{Glover23} have attacked the Tammes problem applying an algorithm based on an iterated dynamic neighborhood search and improving several results in the literature for the Tammes problem. 

With this in mind we have also implemented a search based on basin hopping for packing on a spherical cap (alongside
to the algorithms that we have described in our previous papers~\cite{Amore23a,Amore23b,Amore23c}) to improve our chances of finding good maxima of the packing fraction. 
We have used the implementation of basin hopping in scipy~\cite{scipy}, taking into account that  the maximization of the packing fraction  needs to be performed indirectly because the algorithm itself minimizes an energy functional.

This approach consists of three stages:
\begin{itemize}
\item in the first stage we generate a random ansatz and evolve it from a small $s^{(1)}_i$ (for instance $s^{(1)}_{\rm i}=2$) up to a larger value $s^{(1)}_{\rm f}$ (for example $s^{(1)}_{\rm f} \approx 10-20$);  the evolution is done using algorithm 1 of ref.~\cite{Amore23a}, with the appropriate modifications required by the problem, and it is typically very fast.

\item in the second stage the energy landscape is explored from $s = s_{\rm i}^{(2)}$ ($s_{\rm i}^{(2)} = s^{(1)}_{\rm f}$)  
up to a larger value $s_{\rm f}^{(2)}$ (for example $s_{\rm f}^{(2)} \approx 500-1000$) using basin hopping; in the simplest case 
a single exploration at $s = s_{\rm i}^{(2)} = s_{\rm f}^{(2)}$ is carried out.  In this exploration the configurations corresponding to local minima of the energy functional are evolved up to a much larger $s$ and the packing fraction of the corresponding configuration is calculated and used in the acceptance criteria. This step introduces extra computational  load but it is needed because of the bias that the energy functional introduces in the search, particularly when $s$ is not large enough: as a matter of fact, typically the densest configurations  obtained by basin hopping (i.e. for $s_{\rm i}^{(2)} \leq s \leq s_{\rm f}^{(2)}$) do not evolve into the densest configurations for $s \gg 1$. On the other hand, the possible solution of carrying out the landscape exploration at much larger values of $s$ is impractical because in this regime one deals with very large numbers and numerical precision may be affected (bear in mind that the energy functional itself is not fixed a--priori because of the parameter $\lambda$ that needs to be adjusted during the calculation:  for $s\gg 1$ if one explores a too different configuration the round--off errors may become too large if the $\lambda$ used is not appropriate).
Of course basin hopping can be used specifying both step--size and temperature.

\item in the third stage, the configuration obtained with BH is evolved from $s_{\rm i}^{(3)} = s^{(2)}_{\rm f}$  up to a very large value of $s$, $s_{\rm f}^{(3)}$ (typically $s_{\rm f}^{(3)} = 10^{12}$); this evolution may be repeated a number of times, adding a random perturbation (this essentially amounts to algorithm 2 of \cite{Amore23a}); because the configuration obtained from stage 2 is usually already very dense, this "shaking" procedure most likely only refines the configurations, slightly increasing the packing fraction, without producing a new one;
\end{itemize}

The time required to execute the procedures described above depends on different factors, particularly on the number of 
iterations used in the BH algorithm and whether BH is carried out at several values of $s$. In general we have used 
a single application of BH ($s_{\rm i}^{(2)} = s_{\rm f}^{(2)}$), to limit the time of execution. The running time grows considerably
with $N$, making the exploration of very large configurations quite demanding. These aspects are certainly important, but at the moment they are not our main concern because packing on spherical caps has never been studied before and comparison with previous work is impossible  (we don't consider in this argument ref.~\cite{Appelbaum20}, whose results  were obtained under simplifying assumptions and for limited values of $N$). 

It would certainly be interesting in future to explore the performance of different algorithms in this problem, such as the algorithm of ref.~\cite{Glover23}, which has recently allowed to find denser configurations for the Tammes problem.

\section{Numerical results}
\label{sec:res}

We have conducted a large number of numerical experiments for spherical caps of five different angular widths, $\theta_{\rm max} = \pi/6$, $\pi/4$, $\pi/2$, $3\pi/4$ and $5\pi/6$. In each of these cases we have looked for the densest configurations, for several values of $N$ 
($2 \leq N \leq 200$ and selected values for larger $N$). Each of these configurations has been obtained using the algorithms described in the previous section and has required a substantial computational work: while we cannot claim to have found the true global maxima of the packing fraction in all these cases due to the challenging nature of the problem (as we have mentioned earlier a number of new candidates for global maxima  
have been found quite recently  in the range $100 \leq N \leq 200$ in ref.~\cite{Glover23} for the case of the Tammes problem, a problem which had been studied intensively in the literature), the values of the density obtained should be nearly maximal.
 
Unlike for other packing problems (such packing in the square, in the circle, or on the sphere, i.e. the Tammes problem) the only results in the literature available for comparison  in this case are those of the paper by Appelbaum and Weiss, Ref.~\cite{Appelbaum99}, which however was limited
to a hemispherical cap with $N \leq 40$.  In their work, those authors made the simplifying choice of placing the detectors on parallel circles on the surface of the hemisphere, conscious of "a possible deviation from the optimal solution"~\cite{Appelbaum99}. Indeed we have found that because of this choice  the density  of most of the configurations reported in ref.~\cite{Appelbaum99} could be improved.

\begin{figure}[H]
\begin{center}
\bigskip
\includegraphics[width=5cm]{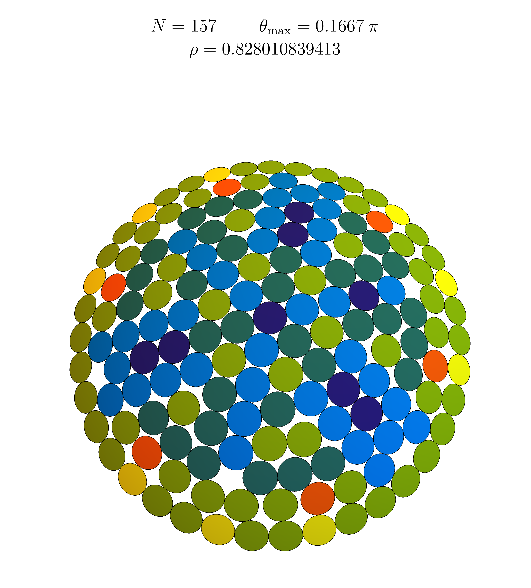}
\includegraphics[width=4cm]{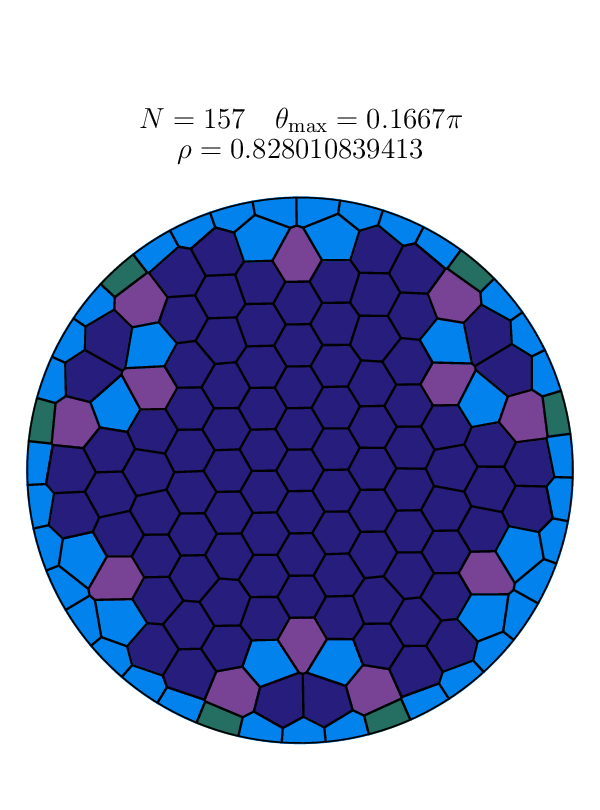}
\bigskip
\caption{Optimal configuration of $N=157$ disks on a spherical cap of width $\theta_{\rm max} \approx \pi/6$.}
\label{Fig_n_157}
\end{center}
\end{figure}

As an example of packing configuration on a spherical cap, in Fig.~\ref{Fig_n_157} we show the case $N=157$ for $\theta_{\rm max} = 0.1667 \approx \pi/6$,
which is invariant under rotations of multiples of $\pi/3$ about the axis passing through the center of the cap. The Voronoi diagram (right plot) shows that the 
topological charge the border is $Q_{border} = 6$, while the internal topological charge adds up to $0$. As a curiosity we observe the small arcs of alternating 
heptagon--pentagon defects, with total charge $-1$ (these arcs are also appearing in some larger configuration, such as $112 \leq N  \leq 118$).
The configuration with $N= 89$ disks, on the other hand, displays a unique closed loop of alternating heptagon-pentagon defects, carrying no topological charge (for reason of space, the figures for these configurations are not included to this paper, but can be accessed at the supplementary material \cite{supp1ementarydata,supp1ementarydata2}).

\begin{figure}[H]
\begin{center}
\bigskip\bigskip\bigskip
\includegraphics[width=10cm]{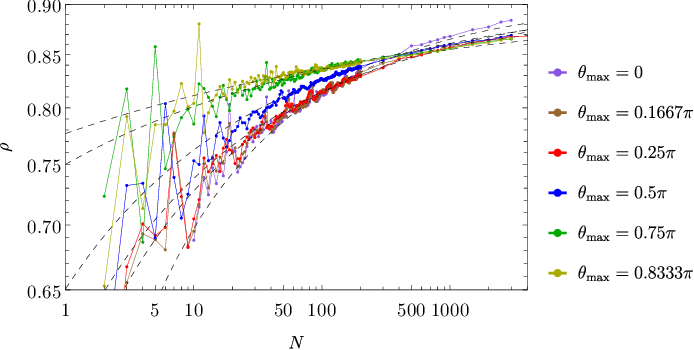}
\bigskip
\caption{Packing fraction for the densest configurations obtained on caps of different width.}
\label{Fig_packing_fraction}
\end{center}
\end{figure}

In fig.~\ref{Fig_packing_fraction} we report the packing fraction for the different cases studied in this paper and include for comparison the
case of packing in a circle, using the values reported by Specht in Packomania~\cite{SpechtRepo}. 
The dashed lines in the plot correspond to the fit
\begin{equation}
\rho(N) = \rho_{\rm plane} - \frac{a}{N^b} 
\label{eq_fit_dens}
\end{equation}
where $\rho_{\rm plane} =  \frac{\pi }{2 \sqrt{3}}$ is the maximal packing fraction in two dimensions.
The values of the parameters $a$ and $b$ for the different $\theta_{\rm max}$ are reported in table \ref{tab:table1}.

\begin{table}[h!]
  \centering
  \caption{Parameters in the fit of eq.~(\ref{eq_fit_dens})}
  \label{tab:table1}
  \begin{tabular}{|c|c|c|}
    \hline
    \textbf{$\theta_{\rm max}$} & \textbf{$a$} & \textbf{$b$} \\
    \hline
    0      & 0.4699       & 0.3505 \\
    0.1667 & 0.3455	      & 0.2851 \\
    0.25   & 0.3098       & 0.2638 \\
    0.5    & 0.2561       & 0.2489 \\
    0.75   & 0.1567       & 0.1712 \\
    0.8333 & 0.1300       & 0.1334 \\
    \hline
  \end{tabular}
\end{table}

Notice that the parameters $a$ and $b$ are both {\sl decreasing with $\theta_{\rm max}$} which results in the particular behavior seen in the figure: at small $N$ the densest packings are obtained with the largest curvature (increasing the curvature helps to increase the internal area compared to the border) while at large $N$ the flatter the domains (i.e. the smallest the curvature) the best, because perfect hexagonal packing can then cover larger regions. 

\begin{figure}[t]
\begin{center}
\bigskip\bigskip\bigskip
\includegraphics[width=10cm]{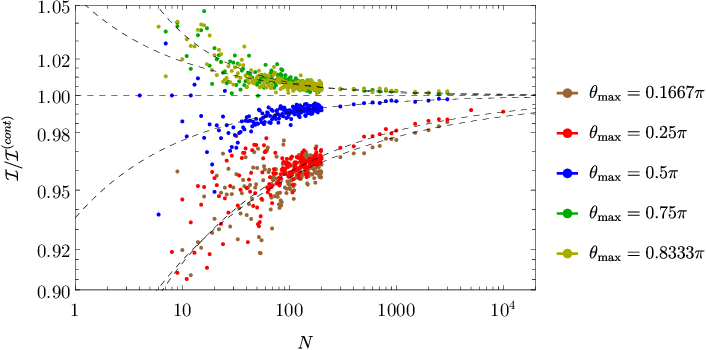}
\bigskip
\caption{Ratio between the discrete and continuum moment of inertia about the $z$ axis for configurations corresponding to different values of $\theta_{\rm max}$. }
\label{Fig_Inertia}
\end{center}
\end{figure}

In the continuum limit ($N \rightarrow \infty$) we expect to reach a uniform distribution of points on the cap: in order to quantify how this limit is 
approached we have plotted in Fig.~\ref{Fig_Inertia}  the ratio between the moments of inertia about the $z$ axis of a discrete and a continuum (uniform) distributions.  The moment of inertia of a uniform spherical cap of width $\theta_{\rm max}$,  $\mathcal{I}^{(cont)}$, is easily obtained to be
\begin{equation}
\mathcal{I}^{(cont)} = \frac{1}{3} + \frac{\cos \theta_{\rm max}}{6} \ .
\end{equation}

The quantity  $\mathcal{I}/\mathcal{I}^{(cont)}$ in the plot can be reasonably described by the behaviour
\begin{equation}
\mathcal{I}/\mathcal{I}^{(cont)} \approx 1 - \frac{\tilde{a}}{N^{\tilde{b}}} \ ,
\end{equation}
where the parameters can be extracted with a fit (the thin dashed lines in the plot are the fits).

The main observation here is that the presence of curvature helps to make the distribution more uniform. 
As a curiosity, we also notice that the cases $N=4,8,12$ on a hemisphere are special because the moment of inertia for the discrete configuration coincides with the continuum value.

An important aspect that needs to be discussed is the nature of the topological defects that appear in the configurations and their dependence on $\theta_{\rm max}$. Euler's theorem of topology, rephrased in term of topological charges, states that the total topological charge 
inside the cap has to equal $6$ (see the discussion in ref.~\cite{Amore23a}), although the way this occurs is not unique and in general will depend 
on $N$ (number of disk) and on $\theta_{\rm max}$.

This analysis can be performed by looking at the Voronoi diagrams of the configurations. To start with, we have used our numerical results 
to look at the total internal topological charge (i.e. the total topological charge carried by internal Voronoi cells) and at the total border topological charge (i.e. the total topological charge carried by border Voronoi cells), $Q^{\rm int}$ and $Q^{\rm border}$ respectively.
Of course one must have $Q^{\rm total} = Q^{\rm int} + Q^{\rm border} = 6$, because of Euler's theorem.
Any interested reader will find all the Voronoi diagrams for the configurations that we have obtained in this study in the supplementary material that accompanies this paper. 

\begin{figure}[H]
\begin{center}
\bigskip\bigskip\bigskip
\includegraphics[width=10cm]{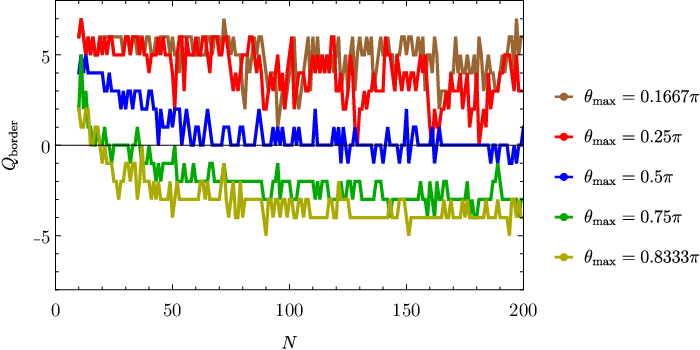}
\bigskip
\caption{Border topological charge for  different values of $\theta_{\rm max}$ as function of $N$. }
\label{Fig_Qborder}
\end{center}
\end{figure}

In Fig.~\ref{Fig_Qborder} we plot the total topological charge on the border as function of $N$, for different values of $\theta_{\rm max}$. The main 
aspect in this plot is that the border topological charge tends to be positive (negative) when the curvature is small (large), while essentially vanishing for the case of the hemisphere.

\begin{figure}[H]
\begin{center}
\bigskip
\includegraphics[width=8cm]{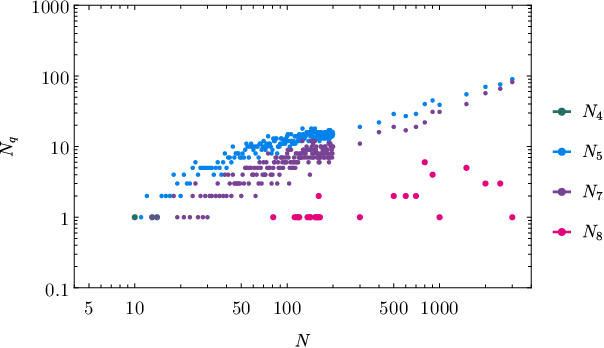} \\
\bigskip
\includegraphics[width=8cm]{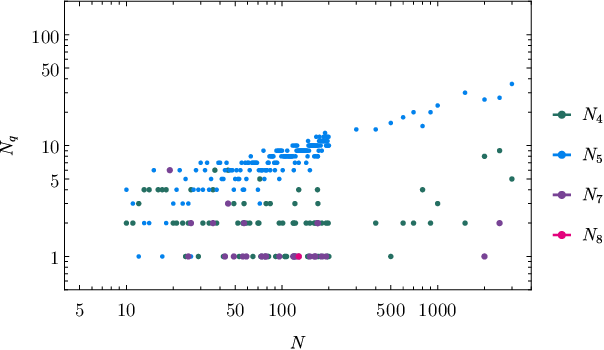}
\bigskip
\caption{Number of internal (upper plot) and external (lower plot) Voronoi cells with $q$ sides for $\theta_{\rm max} = 0.5 \pi$ as function of $N$. }
\label{Fig_N_05}
\end{center}
\end{figure}

\begin{figure}[H]
\begin{center}
\bigskip\bigskip\bigskip
\includegraphics[width=8cm]{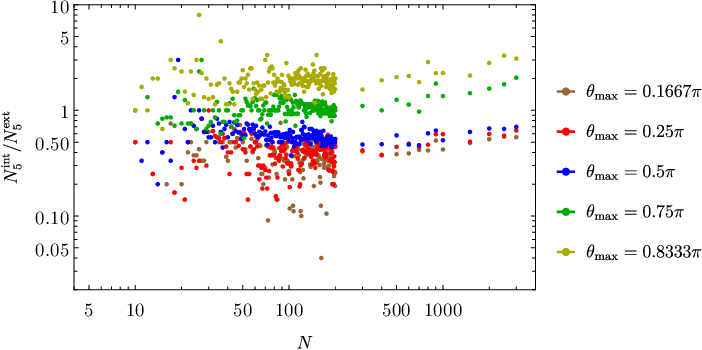}
\caption{Ratio between the number of internal and external pentagonal Voronoi cells  as function of $N$. }
\label{Fig_N5ratio}
\end{center}
\end{figure}

In Fig.~\ref{Fig_N_05} we plot the number of internal (left plot) and external (right plot) Voronoi cells with $q$ sides for $\theta_{\rm max} = 0.5 \pi$ as function of $N$. Similar plots for the other values of $\theta_{\rm max}$ are qualitatively similar.  To better appreciate the specific features 
that arise when changing the curvature, we plot in  Fig.~\ref{Fig_N5ratio} the ratio  between the number of internal and external pentagonal Voronoi cells. This ratio does not display a strong dependence of $N$, but it is seen to increase as $\theta_{\rm max}$ grows (this can be understood in terms of the shrinkage of the border for $\theta_{\rm max} \rightarrow \pi$).

\begin{figure}[H]
\begin{center}
\bigskip
\includegraphics[width=8cm]{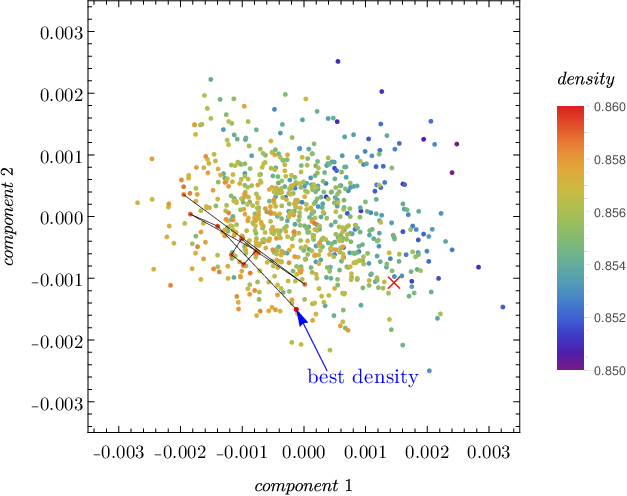}
\bigskip
\caption{Principal component analysis for packing configurations of $1000$  disks on a spherical cap of width $\theta_{\rm max} = 0.5 \pi$. The red cross in the plot marks the position of the configuration $49$, which is the most similar to the densest configuration found among the ones obtained.}
\label{Fig_PCA}
\end{center}
\end{figure}

\begin{figure}[H]
\begin{center}
\bigskip
\includegraphics[width=8cm]{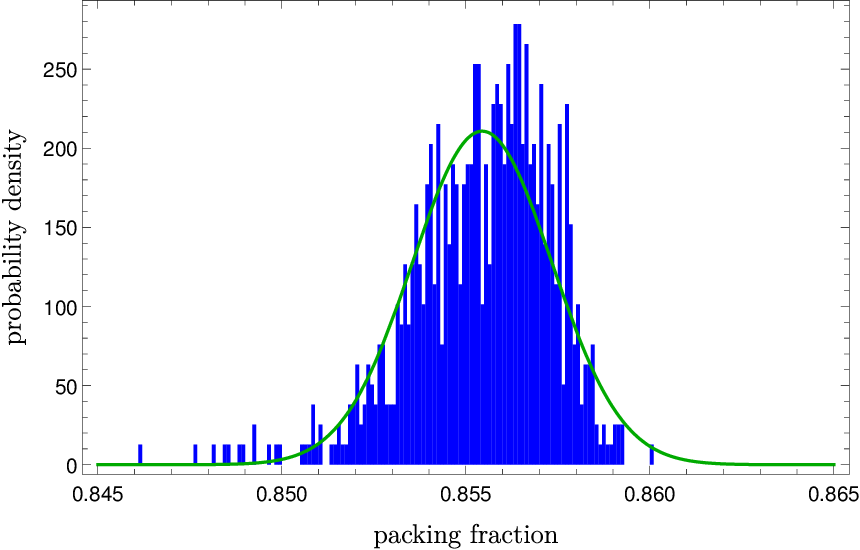}
\bigskip
\caption{Histogram for the packing configurations Fig.~\ref{Fig_PCA}.  }
\label{Fig_hist}
\end{center}
\end{figure}

Another aspect that emerges from looking at the Voronoi diagrams of the configurations (see \cite{supp1ementarydata}) is the fact that 
the defects tend to be located closer to the border at small curvature, while moving to the central region at larger curvature. Additionally, particularly for the case $\theta_{\rm max} = 0.5 \pi$ (but not exclusively), we see in many cases the emergence of chain of defects (pentagon--heptagon) that go from the border to the interior.

In Fig.~\ref{Fig_PCA} we have performed a principal component analysis (PCA) on $800$ independently generated configurations of $1000$ disks on a spherical hemisphere ($\theta_{\rm max} = 0.5 \pi$).  Each point in this plot represents a configuration obtained by numerical optimization, with its color being related to the corresponding value
of the packing fraction (the best density is identified by the arrow in the plot).  The thin line in the figure
connects the first few densest configurations found: one should appreciate that these configurations are typically well separated.

Here PCA gives us an approximate idea of the density landscape associated to the problem. The points in these plot correspond to configurations 
which are more probable to obtain with a number of statistically independent experiments: when using a global optimization algorithm, such as basin--hopping, 
the exploration of the landscape relies, among other things, on the use of temperature as a parameter to control the acceptance/rejection of non-optimal solutions and allows not to get stuck in a small region of the landscape, particularly in the preliminary phases of the algorithm.

Notice also that the probability of finding a configuration with a given packing fraction by means of independent trials appears to follow approximately a gaussian distribution, as seen in Fig.~\ref{Fig_hist}.

The points in Fig.~\ref{Fig_PCA} represent packing configurations on a spherical cap, with the same $N$ and $\theta_{\rm max}$: we assume that
these configurations are arranged in order of decreasing density, $\left\{ \mathcal{C}_0 , \mathcal{C}_1, \dots  \right\}$, 
such that $\rho_0 \geq \rho_1 \geq \dots$.

We want to quantify the similarity between different configurations, $\mathcal{C}_a$ and $\mathcal{C}_b$, represented by the cartesian coordinates ${\bf r}_i^{(a)}$ and ${\bf r}_i^{(b)}$ ($i=1,\dots,N$) respectively. Since any packing configuration can be rotated by an arbitrary angle about the vertical axis, we rotate one of the configurations, say $\mathcal{C}_b$, of an angle $\xi$  and call ${\bf r}_i^{(b)}(\xi)$ the resulting coordinates.

The quantity
\begin{equation}
\mathcal{A}_{ab}(\xi) \equiv \sum_{i=1}^N \min \left\{ \left| {\bf r}_i^{(a)} - {\bf r}_1^{(b)}(\xi)\right|^2, \left| {\bf r}_i^{(a)} - {\bf r}_2^{(b)}(\xi)\right|^2,\dots, \left| {\bf r}_i^{(a)} - {\bf r}_N^{(b)}(\xi)\right|^2 \right\}
\end{equation}
measures the degree of superposition between the two configurations at a specific angle $\xi$ (in particular if $\mathcal{A}(\xi)=0$ for some $\xi$, then the two configurations coincide). 

$\mathcal{A}_{ab}(\xi)$ is a periodic function $\xi$ with period $2\pi$ and one should look for the value of $\xi$ which corresponds to the global minimum of $\mathcal{A}_{ab}$. We call $\xi_{\rm min}$ this value and $\mathcal{A}_{ab}^{(min)} \equiv \mathcal{A}_{ab}(\xi_{\rm min})$, which provides then a direct indication on the similarity between the two configurations (an analogous quantity can be defined  by considering only the points on the border of the cap).

\begin{figure}[H]
\begin{center}
\bigskip
\includegraphics[width=8cm]{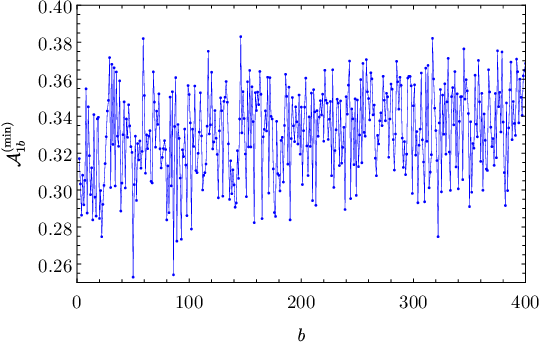}
\bigskip
\caption{$\mathcal{A}_{1b}^{(\rm min)}$ for the the configurations of Fig.~\ref{Fig_PCA}.
The configuration for $b = 50$ is the most similar to the one with the lowest energy among those obtained.
}
\label{Fig_affinity}
\end{center}
\end{figure}

In Fig.~\ref{Fig_affinity} we plot $\mathcal{A}_{1b}^{(\rm min)}$ for the the configurations of Fig.~\ref{Fig_PCA}: remarkably, we see that the configuration which is most similar to the densest configuration that we have found for $N=1000$ disks on the hemisphere is not very close in density (in fact there are $49$ other solutions with larger density).

\section{The fate of CHP}
\label{sec:CHP}

As we have mentioned in the Introduction, the configurations in the circle ($\theta_{\rm max} \rightarrow 0$) that maximize the packing fraction are invariant under rotations of $\pi/3$ about the center at special values of $N$ ($N = 3 k (k+1)+1$  with $k=0,1,\dots, 5$), called {\sl hexagonal numbers}.  Graham and Lubachevski (GL) discovered these configurations in ref.~\cite{Graham97b} and called them {\sl curved hexagonal packing}.

Because their density  for $N \rightarrow \infty$ tends to a finite value smaller than $\pi/\sqrt{12}$, they cease to be optimal at a finite value of $N$: the numerical evidence gathered by Graham and Lubachevski points to $N=91$ ($k=5$) being the last CHP in the circle.  A similar occurrence is observed in regular polygons with a number of sides multiple of $6$, as recently found in ref.~\cite{Amore23c} (in the case of the dodecagon, in particular, it was found that $N=127$ is still CHP). In all these examples, however, the domain (either circle or regular polygon) is flat.

When working on a spherical cap, with a small angular opening $\theta_{\rm max}$, we can be very close to the conditions explored by GL and we expect that the densest configurations up to $N=91$ to be CHP. As the curvature of the domain is increased these configurations evolve adapting 
to the changed domain and eventually are replaced by configurations with no (or different) symmetry. A qualitative argument to explain this behavior is the following: in a CHP the border is fully packed; as the curvature grows, there is more space internally available for packing, which however cannot be exploited by CHP because the size of the disks is already maximal (the border disks cannot swell). After reaching a critical value of 
$\theta_{\rm max}$, suddenly there will be a {\sl disordered} configuration  with larger density which will take advantage of the extra space, by removing outer disks and placing them internally.

We can also describe the behavior of CHP as a function of $\theta_{\rm max}$ {\sl quantitatively}: 
from the requirement that $6k$ disks are fully packing the border one obtains the transcendental condition
\begin{equation}
\theta_{\rm min} =  \sin \left(\frac{\pi }{6 k}\right) \sin (\theta_{\rm max} -\theta_{\rm min}) \ , 
\label{eq_dens_CHP}
\end{equation}
which, for $\theta_{\rm max} \rightarrow 0$, has the solution
\begin{equation}
\theta_{\rm min} = \theta_{\rm max} \frac{\sin \left(\frac{\pi }{6 k}\right)}{\sin \left(\frac{\pi }{6 k}\right)+1} \ .
\end{equation}

Taking the limit  $\theta_{\rm max} \rightarrow 0$, the packing fraction reduces to
\begin{equation}
\lim_{\theta_{\rm max}\rightarrow 0} \rho = (3 k (k+1)+1) \left(\frac{1}{\sin \left(\frac{\pi }{6 k}\right)+1}-1\right)^2 \ ,
\end{equation}
which corresponds to the expression given in eq.(4) of ref.~\cite{Graham97b} for CHP in the circle.

For $k \rightarrow \infty$ (which implies $N\rightarrow \infty$) one obtains 
\begin{equation}
\lim_{k \rightarrow \infty} \lim_{\theta_{\rm max}\rightarrow 0} \rho = \frac{\pi^2}{12} < \rho_{\rm plane} \equiv \frac{\pi}{\sqrt{12}} \ ,
\end{equation}
where $\rho_{\rm plane}$ is the packing fraction obtained with perfect hexagonal packing on the infinite plane.

\begin{figure}[htbp]
\begin{minipage}[b]{\linewidth}
\centering
\includegraphics[width=7cm]{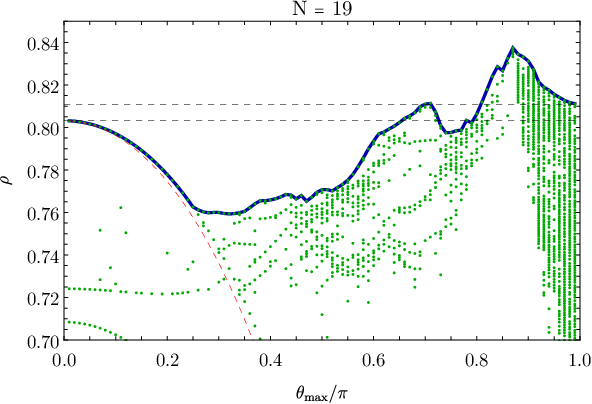} 
\caption{Packing fraction for $N=19$ as a function of $\theta_{\rm max}$ (solid line). The green dots correspond to the different configurations found in $500$ runs. The dashed horizontal lines correspond to the values of the packing fraction for $\theta_{\rm max} = 0.01 \pi$ and $\theta_{\rm max} = 0.99 \pi$. The red dashed line correspond to the density obtained by numerically solving eq.~(\ref{eq_dens_CHP})}
\label{Fig_19_range}
\end{minipage}
\begin{minipage}[b]{\linewidth}
\centering
\includegraphics[width=4.5cm]{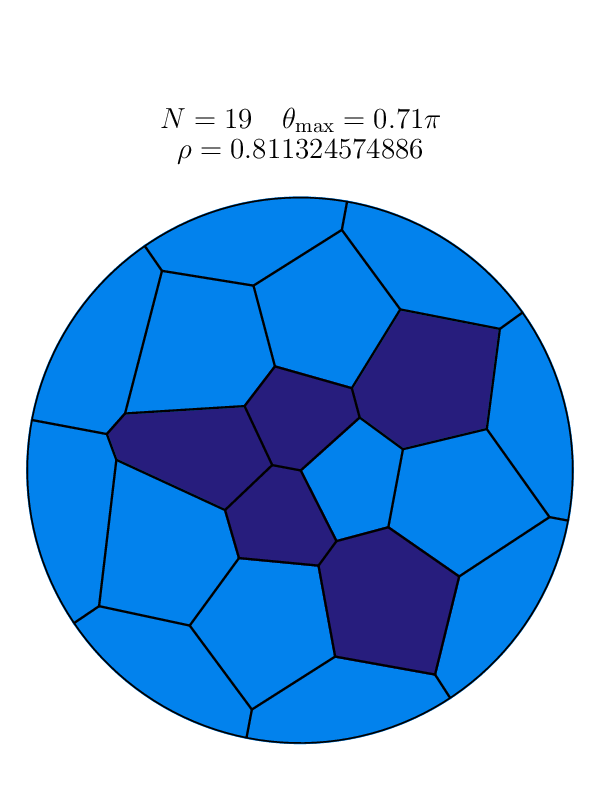} \hspace{1cm}
\includegraphics[width=4.5cm]{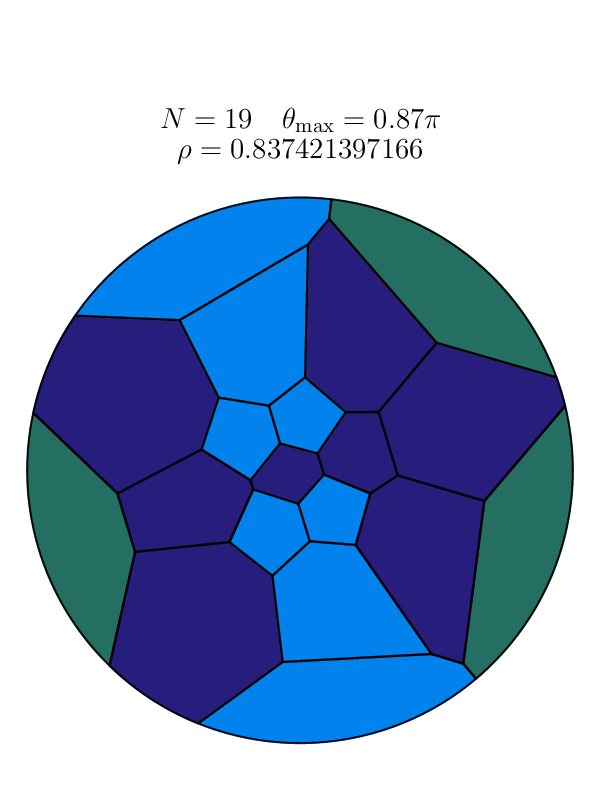}
\caption{Voronoi diagrams for the configurations of $19$ disks corresponding to a maximum of the packing fraction in Fig.~\ref{Fig_19_range}.}
\label{Fig_19_maxdens}
  \end{minipage}
\end{figure}

In Fig.~\ref{Fig_19_range} we plot the packing fraction for $N=19$ as a function of $\theta_{\rm max}$. 
There are several  aspects of interest in this figure: to start with, the packing fraction starts
decreasing around $\theta_{\rm max} \approx 0.01 \pi$ ($\rho \approx 0.8031$) until reaching a plateau around $\theta_{\rm max} \approx 0.26 \pi$ (this behavior is well described by the packing fraction obtained using $\theta_{\rm min}$ from eq.~(\ref{eq_dens_CHP}), represented by the red dashed line in the plot).  For $\theta_{\rm max} \geq 0.26 \pi$, the packing fraction first stabilizes and then starts increasing in value, reaching a first maximum at $\theta_{\rm max} \approx 0.71 \pi$, which is just slightly denser that the original CHP. A second larger maximum is then reached at $\theta_{\rm max} \approx 0.87$, with a packing fraction of $\rho \approx 0.8345$  (the horizontal dashed lines in the plot correspond to the densities for $\theta_{\rm max} = 0.01 \pi$ and $\theta_{\rm max} = 0.99 \pi$).

The second aspect of interest concerns the number of less dense configurations as $\theta_{\rm max}$ changes: the green points in this plot correspond to all the configurations that were found while running the algorithm (the algorithm runs $500$ times at each value of $\theta_{\rm max}$). We have not made an effort to find {\sl all} such configurations, but the plot is nonetheless indicative. In particular we see that at smaller $\theta_{\rm max}$ there are fewer configurations and these are well separated in density (as a matter of fact a large gap in density separates 
the densest configuration from the next one for $\theta_{\rm max} < 0.25 \pi$). Conversely, there are many configurations in a small range of density at large $\theta_{\rm max}$.

The Voronoi diagrams of the two maxima of the density mentioned earlier are plotted in Fig.~\ref{Fig_19_maxdens}: in this case we see that the  configurations are invariant with respect to reflections about an equatorial plane going through the north pole. Additionally we see that the topological charge carried by the border vanishes in both cases.

\begin{figure}[htbp]
\begin{minipage}[b]{\linewidth}
\centering
\includegraphics[width=7cm]{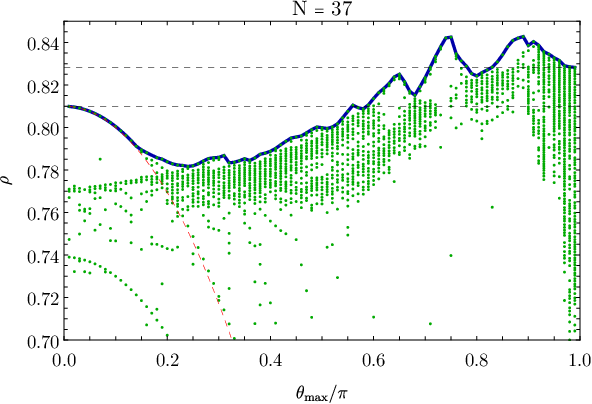}
\bigskip
\caption{Packing fraction for $N=37$ as a function of $\theta_{\rm max}$ (solid line). The green dots correspond to the different configurations found in $500$ runs. The dashed horizontal lines correspond to the values of the packing fraction for $\theta_{\rm max} = 0.01 \pi$ and $\theta_{\rm max} = 0.99 \pi$. The red dashed line correspond to the density obtained by numerically solving eq.~(\ref{eq_dens_CHP})}
\label{Fig_37_range}
\end{minipage}
\begin{minipage}[b]{\linewidth}
\centering
\includegraphics[width=4.5cm]{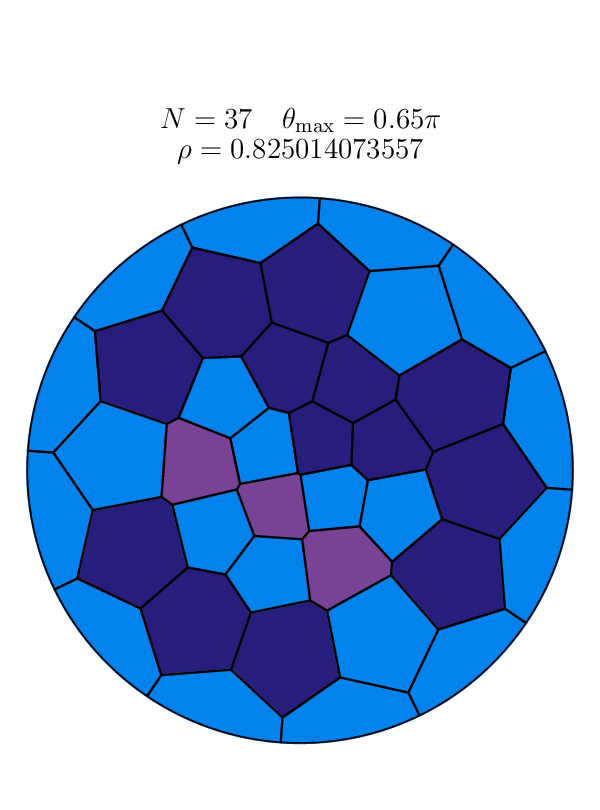} \hspace{1cm}
\includegraphics[width=4.5cm]{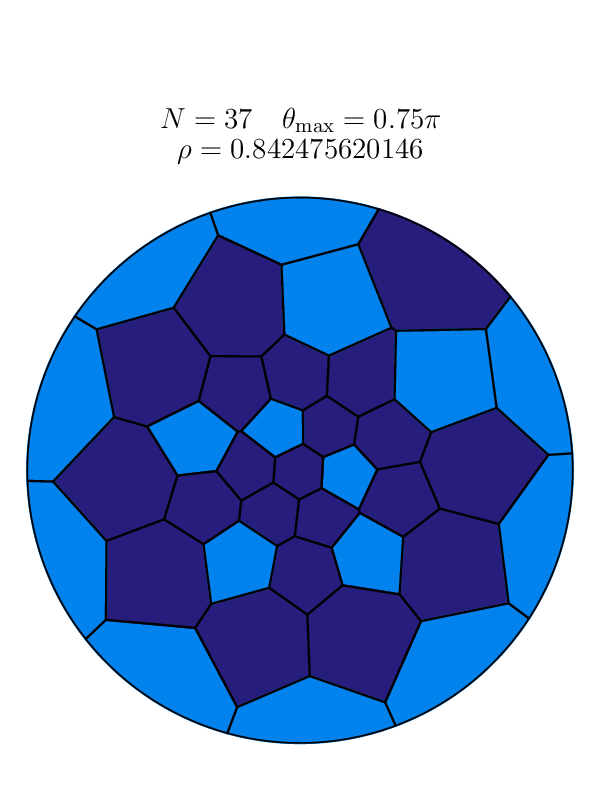} \\
\bigskip
\includegraphics[width=4.5cm]{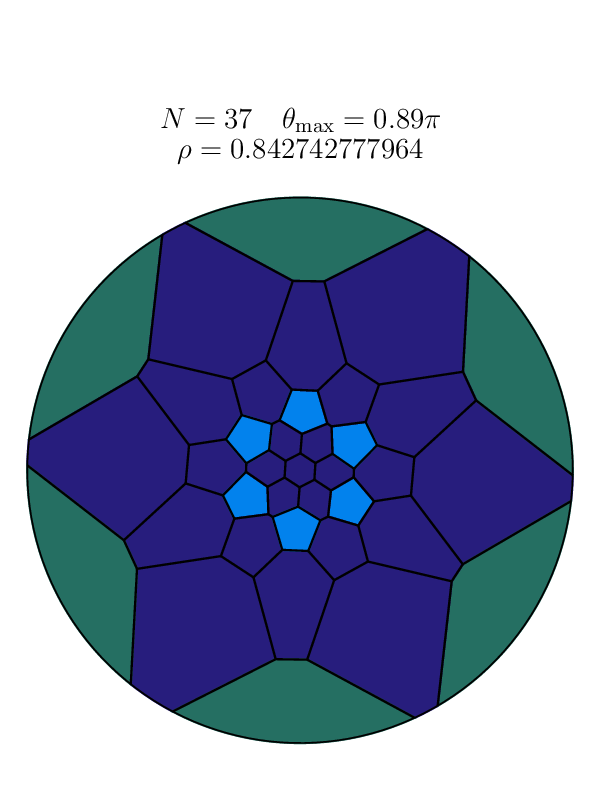} \\
\caption{Voronoi diagrams for the configurations of $37$ disks corresponding to a maximum of the packing fraction in Fig.~\ref{Fig_37_range}.}
\label{Fig_37_maxdens}
\end{minipage}
\end{figure}

The same analysis that we have just made can be repeated for the case of $N=37$, $61$ and $91$. The packing fraction for $N=37$ is plotted in Fig.~\ref{Fig_37_range}. In this case there are three maxima of interest, reported in Fig.~\ref{Fig_37_maxdens}.
Once again, we see that these configurations either possess a symmetry with respect to reflection as in the previous case 
($\theta_{\rm max} =0.65\pi$ and $\theta_{\rm max} =0.75\pi$) or with respect to rotations about the vertical axis (the symmetry axis of the spherical cap) by multiples of $\pi/3$ for $\theta_{\rm max} = 0.89\pi$ (this case is remarkable because it is the same symmetry of the original CHP).

\begin{figure}[htbp]
\begin{minipage}[b]{\linewidth}
\centering
\includegraphics[width=7cm]{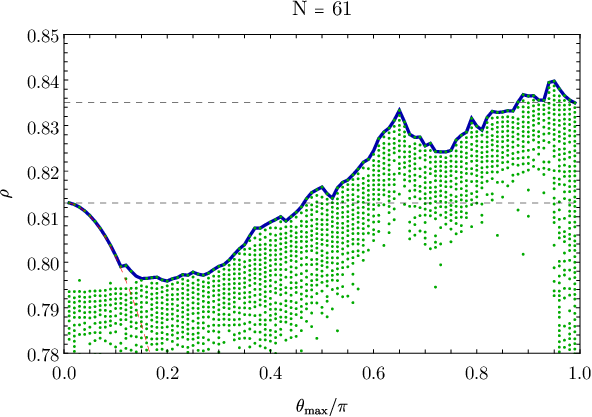}
\bigskip
\caption{Packing fraction for $N=61$ as a function of $\theta_{\rm max}$ (solid line). The green dots correspond to the different configurations found in $500$ runs. The dashed horizontal lines correspond to the values of the packing fraction for $\theta_{\rm max} = 0.01 \pi$ and $\theta_{\rm max} = 0.99 \pi$. The red dashed line correspond to the density obtained by numerically solving eq.~(\ref{eq_dens_CHP}).}
\label{Fig_61_range}
\end{minipage}
\begin{minipage}[b]{\linewidth}
\centering
\includegraphics[width=4.5cm]{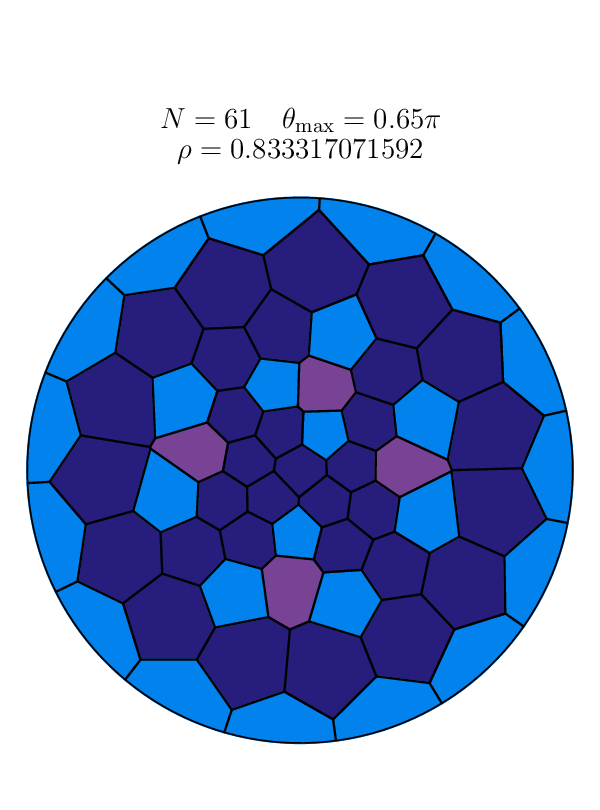} \hspace{1cm}
\includegraphics[width=4.5cm]{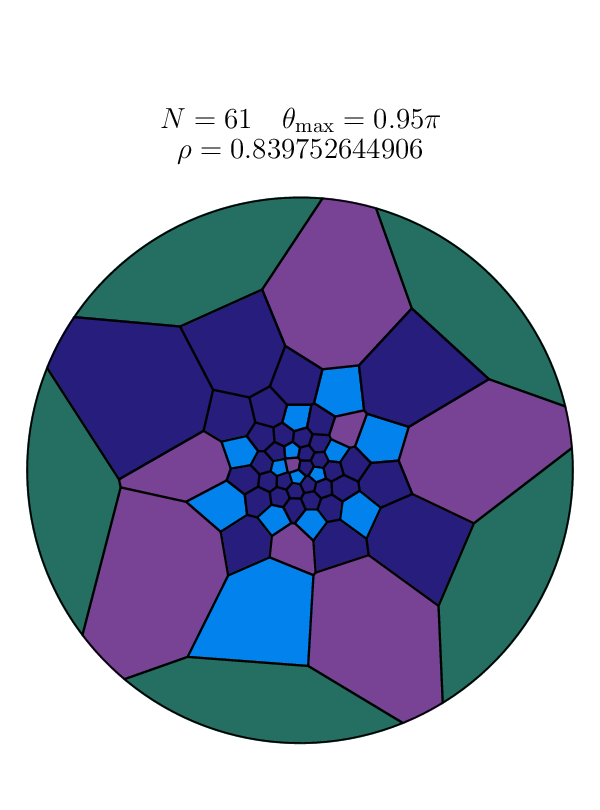} \\
\caption{Voronoi diagrams for the configurations of $61$ disks corresponding to a maximum of the packing fraction in Fig.~\ref{Fig_61_range}.}
\label{Fig_61_maxdens}
\end{minipage}
\end{figure}

\begin{figure}[htbp]
\begin{minipage}[b]{\linewidth}
\centering
\includegraphics[width=7cm]{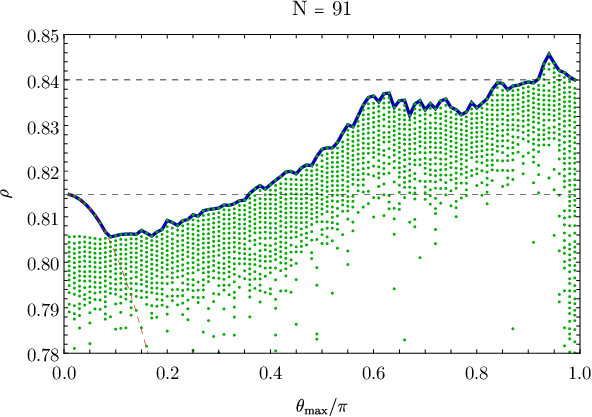}
\bigskip
\caption{Packing fraction for $N=91$ as a function of $\theta_{\rm max}$ (solid line). The green dots correspond to the different configurations found in $500$ runs. The dashed horizontal lines correspond to the values of the packing fraction for $\theta_{\rm max} = 0.01 \pi$ and $\theta_{\rm max} = 0.99 \pi$. The red dashed line correspond to the density obtained by numerically solving eq.~(\ref{eq_dens_CHP}).}
\label{Fig_91_range}
\end{minipage}
\begin{minipage}[b]{\linewidth}
\centering
\includegraphics[width=4.5cm]{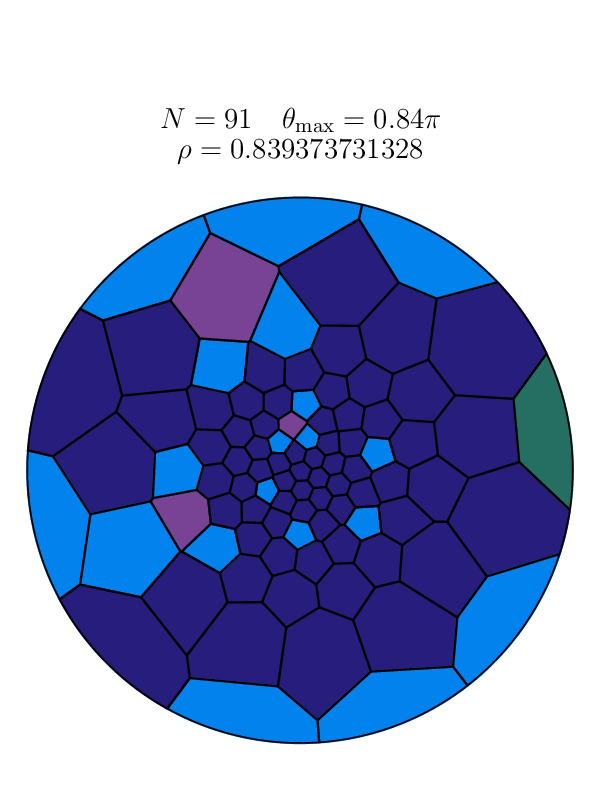} \hspace{1cm}
\includegraphics[width=4.5cm]{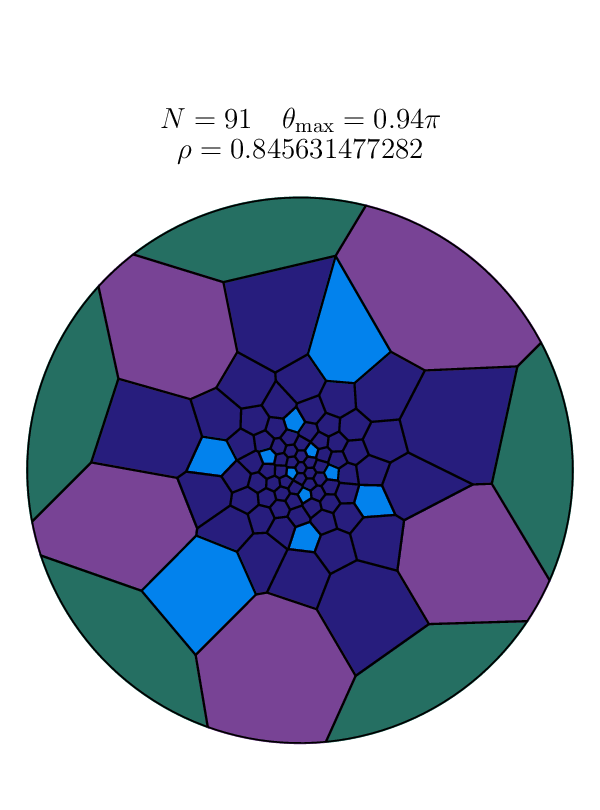} \\
\caption{Voronoi diagrams for the configurations of $91$ disks corresponding to a maximum of the packing fraction in Fig.~\ref{Fig_91_range}.}
\label{Fig_91_maxdens}
\end{minipage}
\end{figure}

\begin{figure}
\begin{center}
\bigskip
\includegraphics[width=7cm]{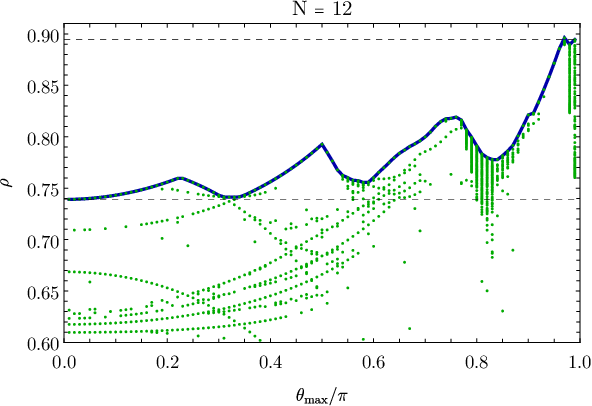}
\bigskip
\caption{Packing fraction for $N=12$ as a function of $\theta_{\rm max}$ (solid line). The green dots correspond to the different configurations found in $500$ runs. The dashed horizontal lines correspond to the values of the packing fraction for $\theta_{\rm max} = 0.01 \pi$ and $\theta_{\rm max} = 0.99 \pi$. }
\label{Fig_12_range}
\end{center}
\end{figure}

\begin{figure}[htbp]
\begin{minipage}[b]{\linewidth}
\centering
\includegraphics[width=7cm]{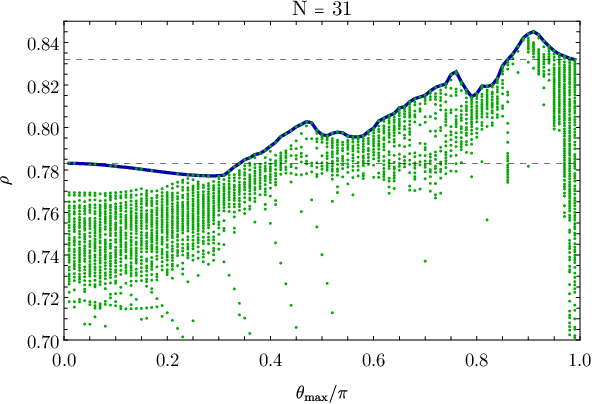}
\bigskip
\caption{Packing fraction for $N=31$ as a function of $\theta_{\rm max}$ (solid line). The green dots correspond to the different configurations found in $500$ runs. The dashed horizontal lines correspond to the values of the packing fraction for $\theta_{\rm max} = 0.01 \pi$ and $\theta_{\rm max} = 0.99 \pi$. }
\label{Fig_31_range}
\end{minipage}
\begin{minipage}[b]{\linewidth}
\centering
\end{minipage}
\end{figure}

{The analogous cases of $N=61$  and $N=91$ are plotted in the Figs.~\ref{Fig_61_range}, \ref{Fig_61_maxdens},  \ref{Fig_91_range}, \ref{Fig_91_maxdens}.
Similarly the packing density for $N=12$ and $N=31$ (non CHP configurations) are plotted  in Figs. \ref{Fig_12_range} and \ref{Fig_31_range}.}

We make an additional note on the presence of symmetrical configurations at $N= 31$, $55$, $85$ for $\theta_{\rm max} = 0.1666 \pi$ 
(for $\theta_{\rm max} = 0.25 \pi$ only the first two cases): analogous configurations where spotted in \cite{Amore23c} for packing inside a regular dodecagon and are particular because they allow a hexagonal packing of the internal disks, by underpopulating the border. Actually, it is easy to realize that they occur
at $N = N_{\rm hex} -6$, where $N_{\rm hex} = 3 k (k+1) +1$, with $k=1,2,\dots$ are the {\sl hexagonal numbers}. Looking at \cite{SpechtRepo} we see 
that these configurations indeed exist also for the circle for $N=31$,$55$,$85$ and $121$.

\section{Conclusions}
\label{sec:concl}

In this paper we have presented the first treatment of circle packing on a spherical cap, a problem that had not been studied before and which is relevant in the description of physical (colloidal suspensions), biological (compound eyes of insects) systems, with potential technological applications (hemispherical cameras, geodesic domes, etc.). 

By using the method of ref.~\cite{Amore23d}, which relies on the introduction of extra degrees of freedom in the problem ("image disks"), it has been possible to circumvent the shortcomings of the packing algorithms of refs.~\cite{Nurmela97,Amore23a,Amore23b,Amore23c}, which only work on a limited class of domains.

We have carried out extensive numerical experiments, obtaining results for caps of different angular width and for different numbers of disks. The analysis of these results has allowed us to  establish some of the properties of these systems empirically, such as the behavior of the packing fraction with $N$ and $\theta_{\rm max}$ (the angular width of the cap), the effect of the curvature on the uniformity of the distribution and on  the topological charge.

We have also discussed the behavior of curved hexagonal packing under the condition of increasing curvature: our numerical results confirm our expectations that CHP should cease to be optimal at some critical value of the curvature, because of the extra internal space made available to the system when curvature is added. Interestingly, the configurations may still reach highly symmetrical local maxima of the packing fraction at some value of the angular width (for instance for $N=37$ at $\theta_{\rm max}=0.89 \ \pi$).

It would be interesting to study in future works circle packing in more complicated geometries (possibly not studied before) or discuss the extension of different algorithms (for instance the one of ref.~\cite{Glover23} or of ref.~\cite{Lai24}, which deals with non-convex multiply connected domains) to the present problem.

\section*{Acknowledgements}
I am grateful to Damian de la Cruz for his help with the Voronoi diagrams and to Prof. Tejeda-Yeomans for useful discussions.
This research is supported by Sistema Nacional de Investigadores (M\'exico).

\section*{Data availability}
The data that support the findings of this study are openly available at Zenodo,  refs.~\cite{supp1ementarydata, supp1ementarydata2}.


\begin{appendices}

\section{Generating random points on the spherical cap}

We consider a spherical cap of angle $\theta_{max}$. The solid angle covered by the cap is $\Omega_{cap} = 4\pi  \sin^2 \frac{\theta_{max}}{2} $.
Similarly, a point on the cap forming an angle $\theta$ with the symmetry axis will lie on the border of a smaller cap of solid angle 
$\Omega_{\theta} = 4\pi  \sin^2 \frac{\theta}{2}$. Therefore the probability that a randomly generated point on the spherical cap forms an angle smaller 
than $\theta$ is
\begin{equation}
p = \frac{ \sin^2 \frac{\theta}{2}}{ \sin^2 \frac{\theta_{max}}{2}}  
\end{equation}
and consequently
\begin{equation}
\theta =  2 \arcsin\left(\sqrt{p} \sin \left(\frac{\theta_{max}}{2}\right)\right) \ .
\label{eq_theta}
\end{equation}

Recalling the parametrization that we use in this paper, $\theta = \theta_{max} \ \sin^2 t$ (and $\phi = u$), we obtain 
\begin{equation}
t = \arcsin \sqrt{\frac{2}{\theta_{max}} \arcsin \left( \sqrt{p} \sin \left(\frac{\theta_{max}}{2}\right)  \right)} \ .
\label{eq_t}
\end{equation}

A uniform distribution of random points on the cap is produced by generating pairs of uniform random numbers, $0 \leq p \leq 1$ and $0 \leq u \leq 2\pi$, 
and using eq.~(\ref{eq_theta}) or (\ref{eq_t}), depending if one wants $\theta$ or $t$.

\begin{figure}[H]
\begin{center}
\bigskip\bigskip\bigskip
\includegraphics[width=5cm]{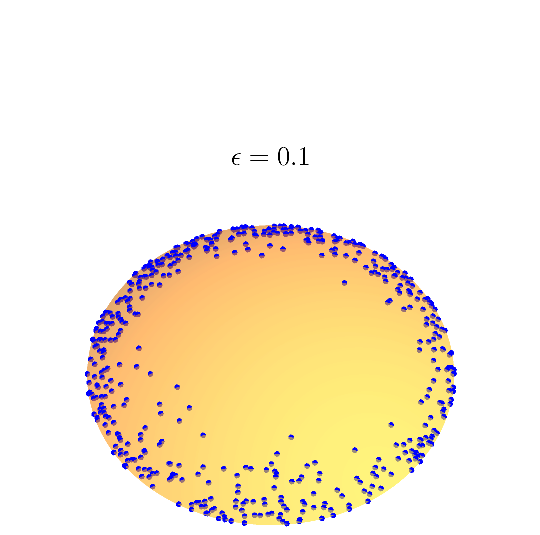} \hspace{1cm}
\includegraphics[width=5cm]{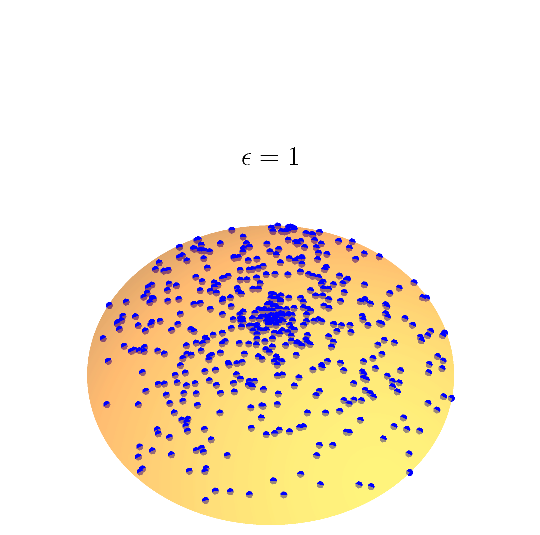}  \hspace{1cm}
\includegraphics[width=5cm]{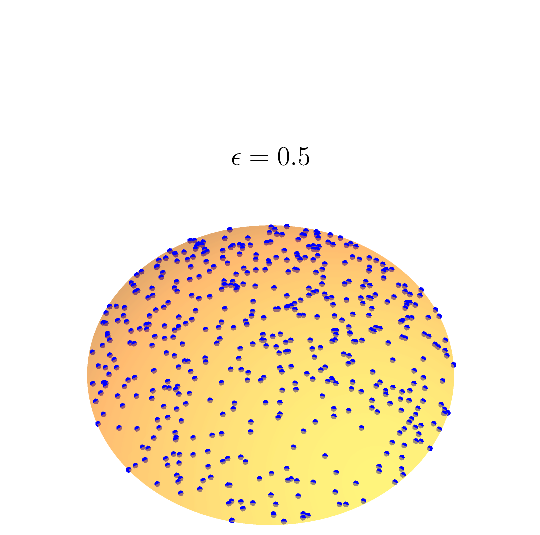}   \\
\caption{Configurations of $500$ points on a spherical cap of angle $\theta_{max} = \pi/8$ generated with eq.~(\ref{eq_theta_eps})}
\label{Fig_random_points}
\end{center}
\end{figure}

More in general, defining
\begin{equation}
\begin{split}
\theta &=  2 \arcsin\left(p^\epsilon  \sin \left(\frac{\theta_{max}}{2}\right)\right) \\
t &= \arcsin \sqrt{\frac{2}{\theta_{max}} \arcsin \left( p^{\epsilon } \sin \left(\frac{\theta_{max}}{2}\right)
   \right)}
\end{split}
\label{eq_theta_eps}
\end{equation}
with $\epsilon >0$,  we will produce a distribution with an excess of points toward the border for $\epsilon<1/2$ and toward the center for $\epsilon >1/2$ (see Fig.~\ref{Fig_random_points}).

We expect that the properties of the initial random configurations to be fed into the algorithm do play a role in facilitating or opposing the search for  
densely packed configurations of disks inside the spherical cap. In a different context, the study of Thomson problem inside a disk that we recently studied in ref.~\cite{Amore23e}, we have observed that the probability of finding the global minimum is very sensitive to the random distribution used for the initial configurations: in that case it was found that the optimal result corresponds to using as probability distribution the (normalized) charge density of the continuum model. On the other hand we found that
using an inappropriate distribution makes the probability of finding the global minimum extremely challenging.

\end{appendices}

\end{document}